\documentclass[epsf]{aa}
\usepackage{natbib}
\usepackage{epsfig}

\begin{document}
\title{Detection of transits of the nearby hot Neptune GJ 436 b}
\subtitle{}
\author{M. Gillon$^{1,  2}$, F. Pont$^1$, B.-O. Demory$^{1,  3}$, F. Mallmann$^3$, M. Mayor$^1$, T. Mazeh$^4$, D. Queloz$^1$, A. Shporer$^4$, S. Udry$^1$, C. Vuissoz$^5$}    

\offprints{michael.gillon@obs.unige.ch}
\institute{$^1$  Observatoire de Gen\`eve, Universit\'e de Gen\`eve, 1290 Sauverny, Switzerland\\
$^2$ Institut d'Astrophysique et de G\'eophysique,  Universit\'e de Li\`ege,  4000 Li\`ege, Belgium \\ $^3$ Observatoire Fran\c{c}ois-Xavier Bagnoud - OFXB, 3961 Saint-Luc, Switzerland\\
$^4$ School of Physics and Astronomy, Raymond and Beverly Sackler Faculty of Exact Sciences, Tel Aviv University, Tel Aviv, Israel\\  
$^5$ Laboratoire d'Astrophysique, Ecole Polytechnique F\'ed\'erale de Lausanne (EPFL), Observatoire, 1290 Sauverny, Switzerland\\}	

\date{Received date / accepted date}
\authorrunning{M. Gillon et al.}
\titlerunning{Detection of transits of the nearby hot Neptune GJ 436 b}
\abstract{This Letter reports on the photometric detection of transits of the Neptune-mass planet orbiting the nearby M-dwarf star GJ 436. It is by far the closest, smallest and least massive transiting planet detected so far. Its mass is slightly larger than Neptune's at $M$ = 22.6 $\pm$ 1.9 $M_\oplus $. The shape and depth of the transit lightcurves show that it is crossing the host star disc near its limb (impact parameter 0.84 $\pm$ 0.03) and that the planet size is comparable to that of Uranus and Neptune, $R$ = 25200 $\pm$ 2200 km = 3.95 $\pm$ 0.35 $R_\oplus$. Its main constituant is therefore very likely to be water ice. If the current planet structure models are correct, an outer layer of H/He constituting up to ten percent in mass is probably needed on top of the ice to account for the observed radius.
 }

\maketitle

\section{Introduction}

While over 200 extrasolar planets have been detected so far, the minority of them that transit their parent stars have the highest impact on our overall understanding of these objects (see review by Charbonneau et al. 2007). They are the only ones with accurate estimates of mass, radius, and, by inference, composition. Further precise monitoring of the brightest of these systems during primary and secondary transits has even permitted the direct study of the planetary atmospheres (e.g. Charbonneau et al. 2002, Marley et al. 2007, Grillmair et al. 2007 ). Until now, this group was composed only of gaseous giant planets\footnote{see http://obswww.unige.ch/$\sim$pont/TRANSITS.htm}, plus the very massive HD 147506 b (Bakos et al. 2007). 

The existence of smaller planets with masses of 5 - 25 $M_\oplus$ was recently uncovered by radial-velocity surveys (e.g. Butler et al. 2004; Udry et al. 2007), raising immediate questions about their constitution. Objects in this range of mass could be composed primarily of H and He gas, water ice, or refractory material (rock/iron). The detection of photometric transits of such an object would bring a preliminary answer to  these questions and has thus been eagerly awaited until now. The photometric precision needed to perform such a detection is beyond the capability of ground-based telescopes for solar-type stars and most of the planetary composition models. However, this is not the case for M dwarfs: their small radius makes possible the detection from the ground of transits of Neptune-sized, or even smaller, planets. Furthermore, existing radial-velocity surveys target relatively bright M-dwarfs, allowing small telescopes to be precise enough to carry out such a transit detection. In this context, we set up a photometric follow-up program of  M-dwarfs known to harbor a close-in low mass planet. 

We report here the first result of our survey, the detection of transits of the Neptune-mass planet orbiting around the nearby M-dwarf star GJ 436 (Butler et al. 2004, hereafter B04; Maness et al. 2007, hereafter M07)  using the 0.6m telescope at the Observatoire Fran\c{c}ois-Xavier Bagnoud (OFXB, Switzerland),  and its confirmation with the 1.2m Euler  telescope (La Silla) and the 1m and 0.46m telescopes at the Wise Observatory (Israel). 

\section{Observations and results}

GJ 436 is a close ($d$ = 10.2 pc)  M2.5V star with $V = 10.67$. It has a low rotation velocity and does not exhibit  particularly strong chromospheric activity nor photometric variability (B04), indicating an age greater than 3 Gyr. 
A periodic Doppler signal revealing the presence of a low mass planetary companion was reported by B04. The new Doppler measurements presented in M07 gave for the planet a minimum mass $M$ sin $i$ =  22.6 $\pm$ 1.9  $M_\oplus$, period, P = 2.64385 $\pm$ 0.00009 days, and indicated an eccentric orbit with $e$ = $0.16 \pm 0.02$. They also revealed a long term trend ($\sim$ 1 m s$^{-1}$ per year), indicating a possible distant companion.

The star was photometrically monitored for transits of its close-in ($a$ = 0.0285 AU) Neptune-mass planet in B04, and the authors concluded that complete transits across the star could be ruled out for gas giant compositions and should be considered as unlikely for solid compositions. We decided nonetheless to include this star in our target list and to observe it from OFXB, judging that the photometric light curve presented in B04 could not completely exclude a shallow and/or grazing transit. 

\subsection{OFXB 0.6m telescope}

The OFXB is a small observatory located in St-Luc, Switzerland, mostly devoted to outreach activities. The telescope is a Newton 0.60m reflector providing a f/3.5 focal ratio. The CCD camera is an Apogee AP47p, equipped with a Marconi 47-10 back-illuminated chip, providing a 20'x20' field of view. This equipment has demonstrated its potential in exoplanet researches by its participation in the characterisation of the transiting planet WASP-2b (Cameron et al., 2007). GJ 436 was monitored during 8 nights over 18 days between April 2nd and April 20th 2007 for a total of 1108 useful exposures. We  scheduled our observations according to the transit windows expected from radial-velocity data, plus some short sequences at random phases to assess the photometric stability of the target. We observed in the $V$-band and defocused to reach 60s exposure time, a good trade-off between time sampling and scintillation mitigation (Gilliland \& Brown, 1992). 

After a standard bias, dark and flatfield correction, all images were reduced with the IRAF/DAOPHOT aperture photometry software (Stetson, 1987), adapting the reduction parameters to the FWHM of each image. Differential photometry was then performed using the flux of nearby stars for which a significant variability could be rejected. The  $rms$ of the OFXB photometry varies from 1.7 to 6 mmag over the different nights. These changes are imputable to differences in weather condition. No stellar variability is seen in phase with the orbital period, a point already quoted in B04. Two clear transit-like events are present in April 2nd and 10th light curves, at the phase expected from the Doppler data. 

\subsection{Wise 1m and  0.46m telescopes}

To secure our tentative detection, we observed GJ 436 at the Wise Observatory (Israel)  on April 24th with two telescopes - 1m and 0.46m - simultaneously. We observed in the $R$-band with the 1m, and used no filter for the 0.46m. The reduction procedure was the same as above. Despite cloudy conditions, we managed to reach a precision high enough to clearly detect  a transit egress at the expected phase with both instruments. 

Figure 1 presents the OFXB and Wise photometry phase-folded using the ephemerides and period presented in M07. A clear transit-like event is visible at phase $\sim$  0.007. 

\begin{figure}
\label{fig:a}
\centering                     
\includegraphics[width=9.0cm]{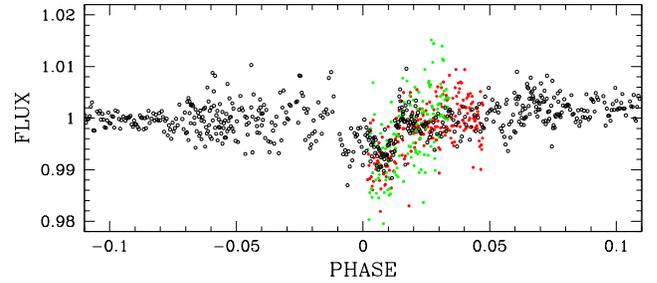}
\caption{ OFXB  (\emph{black}) and Wise (\emph{red: 1m, green: 46cm}) photometry phase-folded using the ephemerids and period presented in Maness et al. (2007).  }
\end{figure}

\subsection{Euler 1.2m telescope}

OFXB and Wise data together confirm the transiting status of GJ 436 b, but do not have high enough quality to firmly constrain the transit parameters. We gathered further observations from the Euler 1.2m  telescope  located at La Silla Observatory (Chile). Observations occured in photometric conditions on May 2nd during 5 hours, encompassing the whole transit window at high airmass (1.8 - 2.1). The same strategy as the one used at OFXB has been applied at Euler during this night ($V$-band filter, 80s exposure time, defocus to $\sim$ 9") resulting in very accurate photometric time series, as can be seen in Fig. 2. The reduction procedure was the same than above. The  $rms$ outside the transit is $\sim$1.2 mmag, while the expected deviation taking into account scintillation and photon noise from the target and the reference stars is $\sim$1 mmag.

\begin{figure}
\label{fig:b}
\centering                     
\includegraphics[width=9.0cm]{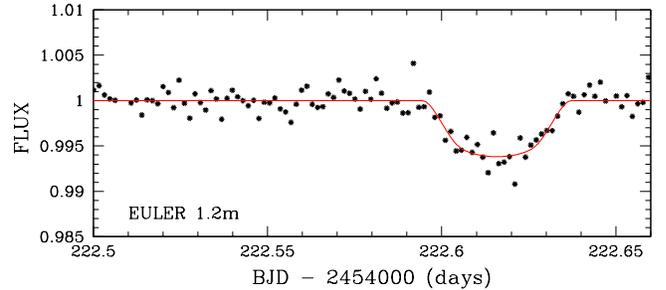}
\caption{Euler V-band transit photometry. The best-fit transit curve is superimposed in red.}
\end{figure}

\section{Parameters of the planet GJ 436 b}

The lightcurves  clearly indicate an almost grazing transit, since its duration is about two times shorter than for a central transit in front of a M2.5V star. The flux drop during the transit is only 0.6\%. These two factors probably explain the non-detection by B04 (these authors state that their data exclude a central transit deeper than 0.4\%).

Since the Euler lightcurve is of a much superior quality than the others, and covers the whole transit, we use only these data for the determination of the parameters. We fitted a transit profile to the Euler data using the Mandel \& Agol (2002) algorithm, the orbital elements in M07 and the quadratic limb darkening coefficients of Claret (2000) for $T_{\rm eff}$ = 3500 K, $\log g$ = 4.5 and [Fe/H] = 0.0 in the $V$ filter. The mass of the parent star was adopted as M = 0.44 $\pm$ 0.04  $M_\odot$  (see discussion in M07). For main-sequence field M dwarfs of such low mass, the mass-radius relation is very tight. Observational constraints from M dwarfs interferometry (see Ribas et al. 2006 and references therein) indicate $M/M_\odot \simeq R/R_\odot$ to within a few percent, and we use this mass-radius relation to set the primary radius. The models of Baraffe et al. (1998) for ages between 1 and 10 Gyr would indicate radius values 0.02 $R_\odot$ lower. On the other hand, the application of the radius calibration from infrared luminosity and temperature by Kervella et al. (2004) indicates a higher value near $R$ = 0.50 $R_\odot$. We thus adopt $M=R=0.44 \pm 0.04$ in solar units for the lightcurve fit. The remaining free parameters are the transit central epoch, the radius ratio and the orbital inclination.

The results of the fit are given in Table 1. The planet crosses the host star disc near its limb (impact parameter 0.84 $\pm$ 0.03). The determination of the orbital inclination lifts the $\sin i$ degeneracy on the planet's mass from the radial velocity orbit, so that $M_{\rm pl}$ = 22.6 $\pm$ 1.9 $M_\oplus$. Our best solution gives a radius of $R_{\rm pl}$ = 25200 $\pm$ 2200 km (3.95 $\pm$ 0.35 $R_\oplus$) for the planet. The uncertainty is mainly due to that on the mass and radius of the primary. The formal uncertainties due to the photon noise of the lightcurve are very small. However, correlated systematics can cause much larger errors on the transit parameters (e.g. Pont, Zucker \& Queloz 2006). Changing the reference stars in the Euler photometry, or using the lower-accuracy data from St-Luc and Wise, leads to changes of $\sim$5\% in the radius ratio.

If the stellar radius is left as a free parameter in the lightcurve fit, the best-fit values are $R=0.46\ R_\odot$, $R_{\rm pl}$ = 26500 km and $i$ = 86$^o$. This is an independent indication that the radius determination of the primary is basically correct. 

\begin{table}
\begin{tabular}{l l } \hline \hline
{\it Star}  & \\
Stellar Mass [$M_\odot$] &  0.44 ($\pm$ 0.04) $^\ast$ \\
Stellar Radius [$R_\odot$] &  0.44 ($\pm$ 0.04)\\
& \\
{\it Planet}  & \\
Period [days] & 2.64385 $\pm$ 0.00009 $^\ast$  \\
Eccentricity &  0.16 $\pm$ 0.02 $^\ast$  \\
Orbital inclination [$^o$] & 86.5 $\pm$ 0.2 \\
Radius ratio &   0.082 $\pm$  0.005  \\
Planet Mass [$M_\oplus$]& 22.6 $\pm$ 1.9 \\
Planet Radius [$R_\oplus$]&  3.95$^{+0.41}_{-0.28}$\\
\ \ \ \ \ \ \ \ \ \ \ \ \ \ \ \ \ \ \ \ [km]&  25200$^{+2600}_{-1800}$\\
$T_{\mathrm tr}$ [BJD] & 2454222.616 $\pm$ 0.001 \\ \hline 
& \\
\end{tabular}
\caption{Parameters for the GJ 436 system, host star and transiting planet. $^\ast$: {\it from M07}.  } 
\label{param}
\end{table}

\section{Discussion}

The measured radius of GJ 436 b is comparable to that of Neptune and Uranus. Figure 3  places it in the context of the mass-radius diagram for Solar System planets and transiting exoplanets. In this part of the mass-radius diagram, the position of a planet is a direct indication of its overall composition, while other factors such as temperature play only a minor role (see e.g. Fortney et al. 2007). In the current paradigm, intermediate-mass planets are composed of some or all of these four layers: an iron/nickel core, a silicate layer, an ice layer (H$_2$O, CH$_4$, NH$_3$), and an H/He envelope. The mass and radius that we measure for GJ 436 b indicate that it is mainly composed of water ice. It is an ``ice giant'' planet like Uranus and Neptune rather than a small-mass gas giant or a very heavy ``super-Earth''. It must have formed at a larger orbital distance, beyond the ``snow line'' where the protoplanetary disc is cool enough for water to condensate, and subsequently migrated inwards to its present orbit.

The temperature profile inside the planet is not expected to modify this qualitative picture. The atmosphere of GJ 436 b must be hot: the equilibrium temperature is 520 K to 620 K depending on the albedo, and a greenhouse effect may heat it to much larger temperatures. Tidal effects from its eccentric orbit must also inject energy in its interior, but the iron, rock and water equations of state are not very sensitive to temperature at high pressure. 

We can ponder whether the planet has an H/He envelope like the ice giants in the Solar System, or if its atmosphere is composed mainly of water vapor. Our best-fit radius value places it slightly above the ``pure ice'' composition mass-radius line of Fortney et al. (2007). A small H/He envelope may thus be required -- even more if an iron/rock core is present as expected. At the upper end of the radius error bar, the H/He envelope would have to represent up to 10\% in mass according to the models of Fortney et al. 2007 (see Figure 3). The lower end of the range is close to the mass-radius line for pure ice planets. Water ice mixed with methane and ammonia is less dense than pure ice under high pressures, so the presence of a significant amount of these compounds within the ice could make the planet large enough despite a rock/iron core to account for the observed radius without invoking an H/He envelope. GJ 436 b could therefore be an ``Ocean Planet'' (L\'eger et al. 2004). Because of the high surface temperature, this would imply a steam atmosphere above supercritical water rather than an Earth-like situation. As methane and ammonia have very low condensation temperatures, this scenario would imply migration from a wide orbit. 
 
\begin{figure}
\label{fig:c}
\centering                     
\includegraphics[width=9.5cm]{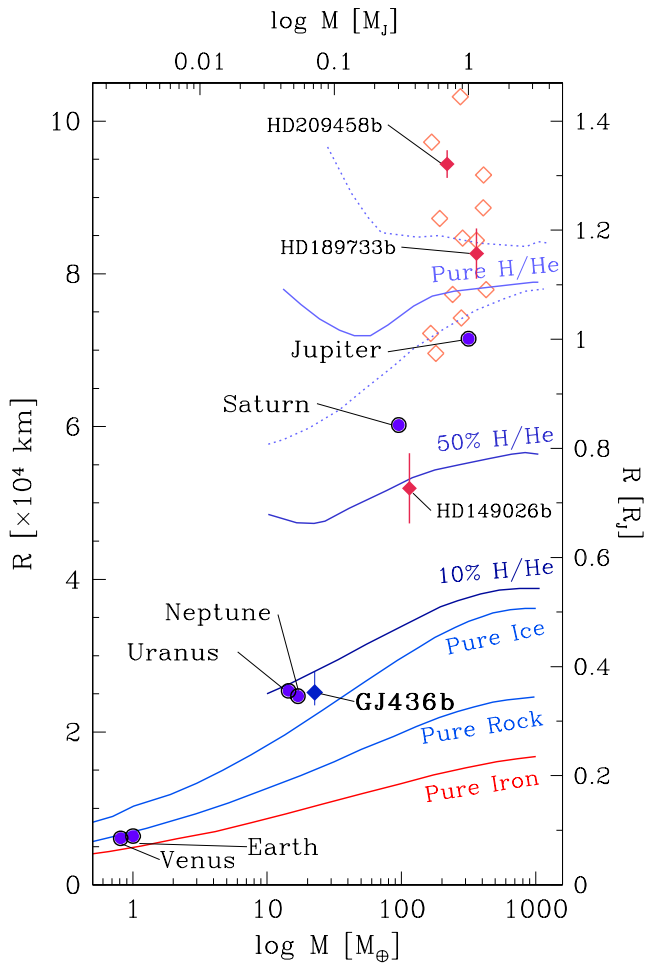}
\caption{Planetary mass-radius diagram (adapted from Fortney et al. 2007) comparing the position of Solar System planets,  transiting hot Jupiters (diamonds), and GJ 436 b. The lines indicate the position of the Fortney et al.  models for different compositions: pure iron, pure silicate, pure water ice (with thermal profiles from Solar System planets), and models for irradiated planets at 0.1 AU from a Solar-type star with a fraction of 10\%, 50\% and 100\% of Hydrogen/Helium. The dotted lines show the models for a cold ($a=10$ AU) and very hot ($a=0.02$ AU) pure H/He gas giant.}
\end{figure}

It could be expected that on such a close orbit, an H/He envelope would quickly evaporate.
But although the planet is very close to its parent star, the small size and low temperature of the primary mean that such an envelope could be retained over long timescales (see M07, Lecavelier 2007).  A more precise radius determination can help determine whether the planet has a water or H/He envelope.

The fact that the orbit of GJ 436 b is not circular indicates a high tidal quality factor $Q$ for the planet, compatible with an ice giant rather than a predominantly rocky planet -- although the eccentricity could be due to the influence of an unseen planetary companion, as pointed out by M07.

GJ 436 b is the first hot Neptune with a radius measurement, and turns out to be a Neptune-like ice giant, mostly composed of water ice, not a rock/iron ``super-Earth'', nor a low-mass gas giant. Its detection illustrates the potential of extensive high-precision photometric follow-up of planets detected by radial velocity. It is the closest transiting planet known, and opens many opportunities for further observations to characterize the planet itself.

\begin{acknowledgements} 
The Fondation de l'Observatoire Fran\c{c}ois-Xavier Bagnoud in St-Luc is greatfully acknowledged for its support. We thank Y. Revaz, E. Ischi, G. Meynet, M. Grenon and J.-C. Pont for their contribution and support to the development of the OFXB. We thank N. Brosch and S. Kaspi for their telescope time at the Wise Observatory telescopes. T.M. was supported by the German-Israeli Foundation for Scientific Research and Development. G. Laughlin is greatfully acknowledged for his inspiring websites transitsearch.org and oklo.org. 

\end{acknowledgements} 

\bibliographystyle{aa}
{}
\end{document}